\documentclass[reprint, twocolumn,
showpacs,
amsmath,amssymb,aps,pra, superscriptaddress]{revtex4-1}

\usepackage{graphicx}
\usepackage{dcolumn}
\usepackage[T1]{fontenc}      
\usepackage{textcomp}

\begin{document}

\title{Tunable source of correlated atom beams}

\author{M.~Bonneau}
\email{bonneau@lens.unifi.it}
\altaffiliation[Present address: ]{INO-CNR, via G. Sansone 1, 50019 Sesto Fiorentino - Firenze, Italy}
\author{J.~Ruaudel}
\author{R.~Lopes}
\author{J.-C.~Jaskula}
\altaffiliation[Present address: ]{Harvard-Smithsonian Center for Astrophysics, Cambridge, Massachusetts 02138, USA}
\author{A.~Aspect}
\author{D.~Boiron}
\author{C.~I.~Westbrook}
\affiliation{Laboratoire Charles Fabry, Institut d'Optique, CNRS, Univ Paris Sud, 2 avenue Augustin Fresnel, 91127 Palaiseau, France}

\begin{abstract}

We use a one-dimensional optical lattice to modify the dispersion relation of atomic matter waves. Four-wave mixing in this situation produces atom pairs in two well defined beams. 
We show that these beams present a narrow momentum correlation, that their momenta are precisely tunable, and that this pair source can be operated in the regimes of low mode occupancy and of high mode occupancy.
\end{abstract} 

\pacs{03.75.Lm, 34.50.Cx, 42.50.Dv, 67.85.Hj}

\maketitle

In quantum optics, existence of mechanisms to produce photon pairs, such as parametric down-conversion, enabled the realisation of several fundamental experiments on quantum mechanics.  
For example, the violation of Bell's inequalities \cite{Aspect1999} or the Hong-Ou-Mandel effect \cite{Hong1987} reveal the surprising properties of quantum correlations in entangled photon pairs. 
These fascinating properties have found applications in quantum information and communications \cite{Pan2012}. 
In analogy to photon pairs, there have been several recent demonstrations of correlated atom pairs production \cite{Greiner2005,Perrin2007,Bucker2011,Lucke2011,RuGway2011,Gross2012,Hamley2012}.  
In particular, momentum correlations of spatially separated samples is an important requirement for the demonstration of an atomic Einstein-Podolsky-Rosen state \cite{Ferris2009,Kofler2012} and the violation of Bell's inequalities.
Such momentum correlations were demonstrated for atom pairs produced by molecule dissociation \cite{Greiner2005} or by spontaneous four-wave mixing in free space through the collision of two Bose-Einstein condensates (BECs) \cite{Perrin2007, Kheruntsyan2012}. 
In these experiments the pairs which were produced lay on a spherical shell. 
This geometry is disadvantageous because many spatial modes are populated, 
and if one wishes to use Bragg diffraction to manipulate and recombine the pairs on a beam splitter \cite{Ferris2009,Kitagawa2011}, 
the vast majority of the pairs is unuseable.

On the other hand, if pair production is concentrated in a small number of modes, 
experimenters can make more efficient use of the generated pairs. 
\begin{figure}[!ht]
\begin{center}
\includegraphics[width=7.5cm]{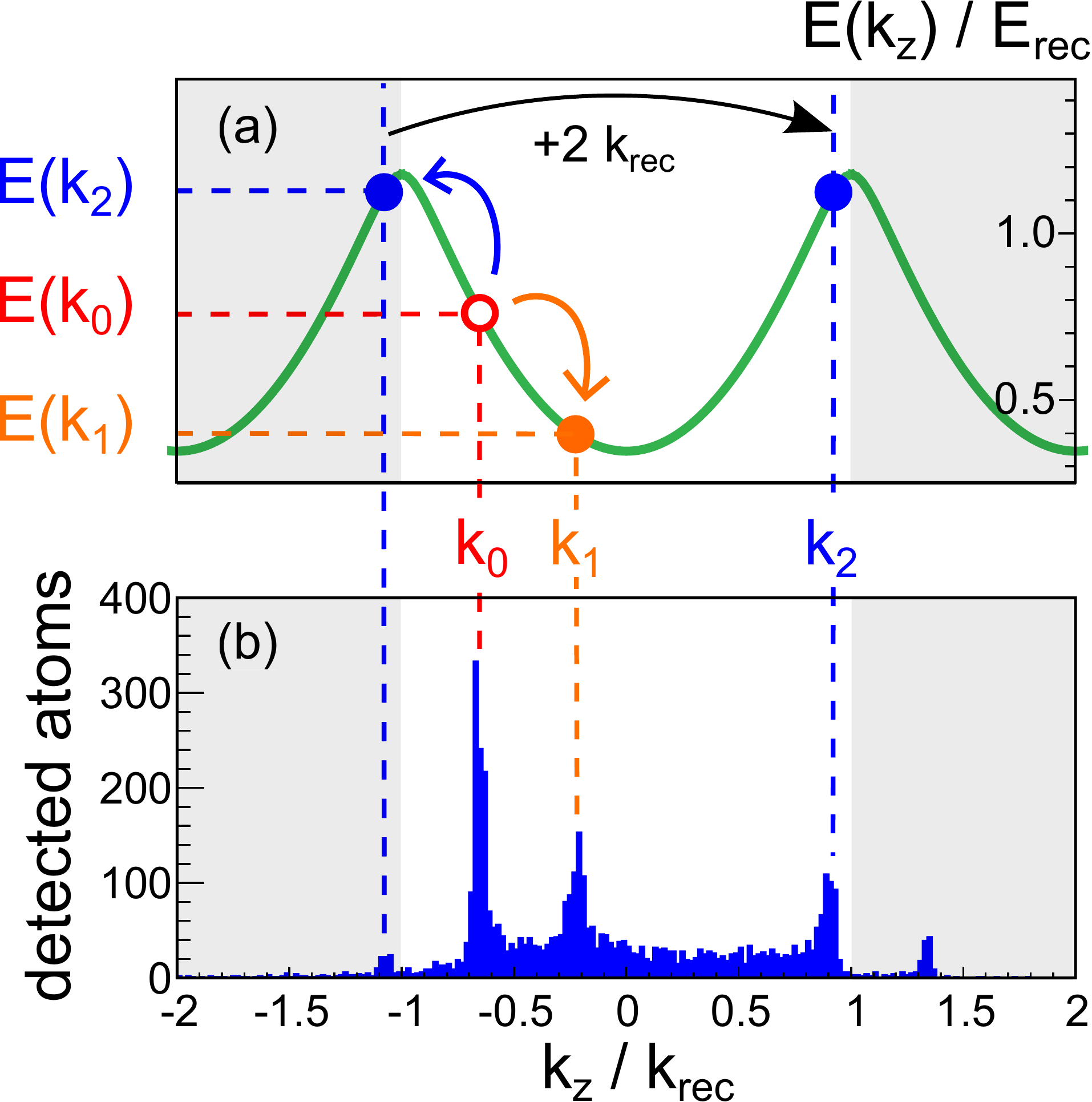}
\caption{\label{princFig}(Color online) (a) 1D pair creation process in an optical lattice with period $\lambda_\textrm{latt.} /2$: The dispersion relation in the first Bloch band (green solid curve) allows scattering of atoms from a BEC with quasimomentum $k_{0}$ (open red circle) in the lattice frame into pairs with quasimomenta $k_{1}$ (filled orange circle) and $k_{2}$ (filled blue circle), so that phase-matching conditions given by energy and momentum conservations are fulfilled. The example here is for a lattice depth ${V_{0}=0.725~E_{\textrm{rec}}}$ and $k_{0}= -0.65~k_{\textrm{rec}}$, with ${k_{\textrm{rec}}=2 \pi / \lambda_\textrm{latt.}}$ the recoil momentum and ${E_{\textrm{rec}}= \hbar^{2} k_{\textrm{rec}}^{2} / 2 m = h \times 44~}$kHz the recoil energy.\\
(b) Vertical single-shot momentum distribution (integrated over the total transverse distribution) measured for these conditions. The three main peaks correspond to the initial BEC and to the macroscopically populated beams centered at $k_{1}$ and $k_{2}$, which are mainly projected in the first Brillouin zone (in white) when the lattice is switched off. As expected, small diffraction peaks at $k_{0}+ 2 k_{\textrm{rec}}$ and $k_{2}- 2 k_{\textrm{rec}}$ are also visible, due to the proximity of $k_{0}$ and $k_{2}$ to the band edge.}
\end{center}
\end{figure}
One can then choose to work either with low mode occupation, the well separated pair regime, or with high mode occupation, referred to as the squeezing regime in Ref.~\cite{Duan2000}. 
An example of twin beams generated in the latter regime is
described in Ref.~\cite{Bucker2011}.
The squeezing regime is well suited to the study of highly entangled multiparticle systems and for investigations of atom interferometry below the standard quantum limit \cite{Bouyer1997, Dunningham2002}. 
The source we study in this Rapid Communication can be operated in both regimes. 
We use atomic four-wave mixing in a one-dimensional (1D) optical lattice, which results in production of atom pairs in two well-defined beams, as proposed in Ref.~\cite{Hilligsoe2005} and demonstrated in Ref.~\cite{Campbell2006}. 
We show that these beams present a narrow momentum correlation, that
their momenta are precisely tunable, and that we can control their intensities. 

In atom optics, four-wave mixing corresponds to scattering into new momentum classes subject to energy and momentum conservation.
In a wave picture, the conservation requirements can be thought of as phase-matching conditions. 
The presence of an optical lattice modifies the free-space atomic dispersion relation, and therefore, for a range of initial quasimomenta $k_0$ \cite{Note1}, the 1D scattering event $2~k_{0} \rightarrow k_{1}+k_{2}$ is allowed, as shown in Fig.~\ref{princFig}(a).
Thus, beginning from a BEC at $k_0$, atom pairs are spontaneously generated along the lattice axis with well-defined quasimomenta $k_1$ and $k_2$. 
We refer to this process as four-wave mixing, but it can also be viewed as a special case of a dynamical instability  \cite{Wu2001,Smerzi2002}, which was studied in the context of coherence \cite{Cataliotti2003,Cristiani2004} and atomic \cite{Fallani2004} losses appearing for a BEC moving in a lattice.

\begin{figure}
\begin{center}
\includegraphics[width=6cm]{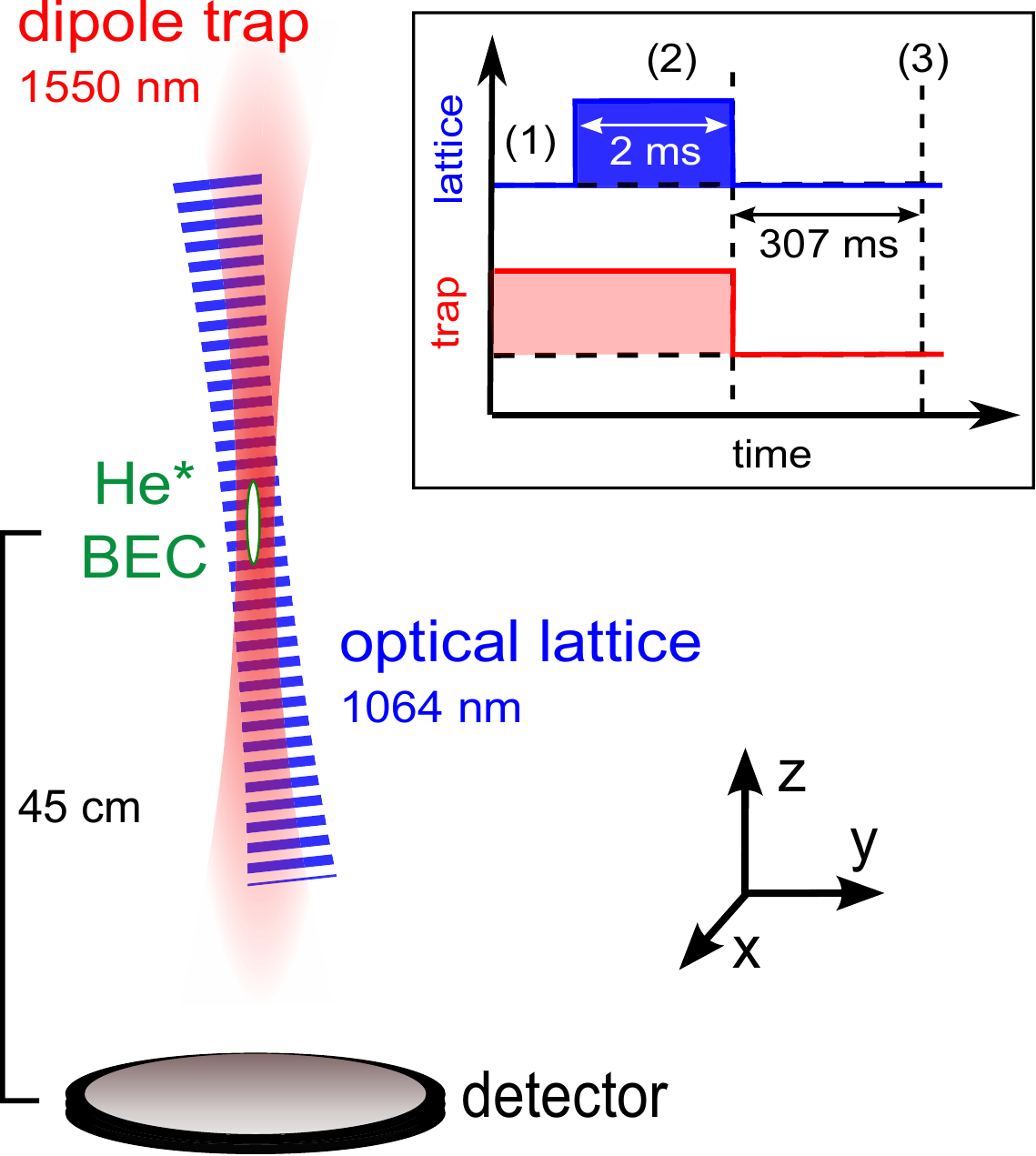}
\caption{\label{manipFig}(Color online) Experimental setup and sequence:\\ 
(1) Initially, a BEC of metastable helium is trapped in a vertical optical potential with a 43~\textmu m waist.\\ 
(2) An optical lattice is suddenly applied in the presence of the trap. 
It is tilted by 7$^{\circ}$ with respect to the trap axis, and is focused on the BEC with a 200~\textmu m waist.\\
(3) After dipole trap and optical lattice switch off, the cloud expands and falls on the 3D resolved single atom detector. 
Given the values of the vertical and transverse Thomas-Fermi radii (0.5~mm and 3~\textmu m), the arrival time and position reflect the 3D momentum distribution, provided the momenta are well above $3 \times 10^{-2}~k_{\textrm{rec}}$ along $z$ and $2 \times 10^{-4}~k_{\textrm{rec}}$ transversely.}
\end{center}
\end{figure}

The experiment is performed on $^{4}\mathrm{He}$ atoms in the $m_{x}=1$ sublevel of the $2\; ^{3}S_{1}$ metastable state. 
The experimental setup and sequence are shown in Fig.~\ref{manipFig}. 
After evaporative cooling in an elongated, vertical dipole trap with frequencies $\nu_{\perp} = 1.5~$kHz and $\nu_{z} = 6.5~$Hz \cite{Partridge2010}, we produce a BEC (or more precisely a quasi-BEC \cite{Petrov2001}) with about $10^{5}$ atoms. 
We then apply a 1D optical lattice with a depth $V_{0}=0.725~E_{\textrm{rec}}$. 
This lattice is tuned 19 nm to the blue of the 1083~nm $2\; ^{3}S_{1}-2\; ^{3}P$ transition of helium. 
It is formed by two counterpropagating $17~$mW beams with $200\,\mu\mathrm{m}$ waists and whose relative detuning $\delta \nu$ can be varied using acousto-optic modulators. We thus control the value of $k_{0}/k_{\textrm{rec}}=h \, \delta \nu/4 E_{\textrm{rec}}$, the BEC's momentum in the lattice frame. 
The lattice is held on for a duration $T_{L}=2~$ms, and suddenly switched off, simultaneously with the optical trap. 
To avoid magnetic perturbation of the cloud during freefall, we apply an RF pulse that transfers $50~\%$ of the atoms to the field insensitive $m_{x}=0$ sublevel \cite{Partridge2010}. 
The atoms remaining in $m_{x}=1$ are subsequently removed by a strong magnetic gradient. 
After a 307~ms mean time of flight, the $m_{x}=0$ atoms fall on a microchannel plate detector, which permits 3D reconstruction of the atomic cloud \cite{Schellekens2005}.\par

As shown in Fig.~\ref{princFig}(b), we observe three main density peaks after the time of flight.
The tallest is the initial BEC. 
The two others are formed by atoms scattered into momentum classes centered in $k_{1}$ and $k_{2}$, whose values are consistent with those expected from the phase-matching 
conditions illustrated in Fig.~\ref{princFig}(a). 
Since the optical lattice is switched off abruptly, the Bloch states of momenta $k_{0}$, $k_{1}$ and $k_{2}$ are projected onto plane waves, mainly in the first Brillouin zone due to the low lattice depth. 
Each of the beams at $k_{1}$ and $k_{2}$ contains about $10^{2}$ detected atoms, which we estimate to correspond to about $2 \times 10^{3}$ atoms per beam. 
We also detect some atoms between the beams, which result from scattering into excited transverse modes \cite{Modugno2004}. 
Due to the low overlap between the transversely excited states and the initial wave function, this transverse excitation is far less efficient than the previously described 1D process. In addition, scattered atoms can also undergo secondary scattering contributing to the background between the beams.\par

\begin{figure}
\begin{center}
\includegraphics[width=8cm]{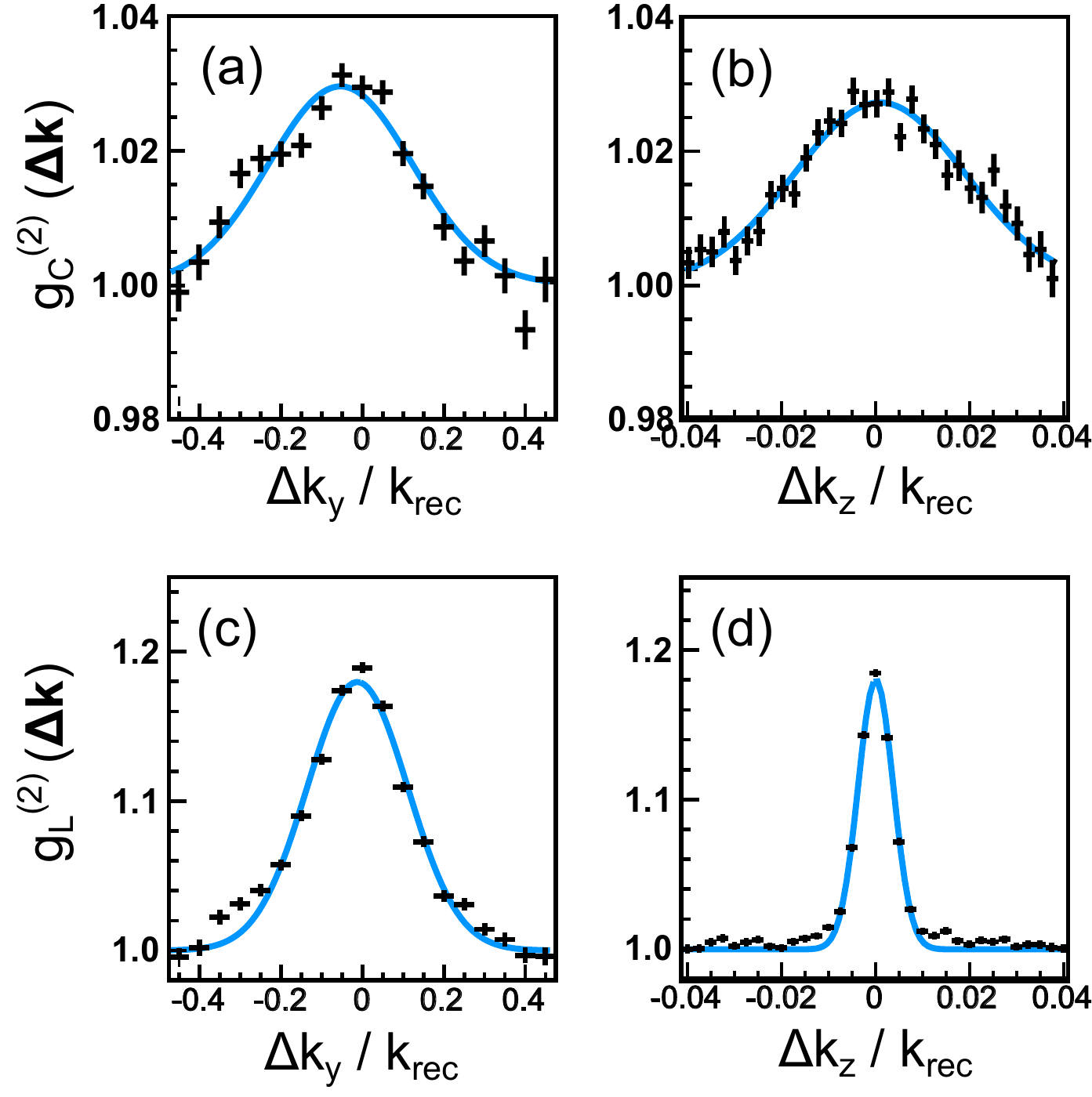}
\caption{\label{corFig}(a,b) Cuts along ($y$,$z$) of the integrated, normalized cross correlation function of the two beams, $g^{(2)}_{C}(\mathbf{\Delta k} ) = \int \mathbf{dk_{i}} \, g_{C}^{(2)}(\mathbf{k_{i}}, \mathbf{k_{j}}+ \mathbf{\Delta k})$. The integration over the momentum distribution $\mathbf{k_{i}}$ is performed on a box with dimensions $L_{k_{x}}=L_{k_{y}}=0.4~k_{\textrm{rec}}$ and $L_{k_{z}}=5 \times 10^{-2}~k_{\textrm{rec}}$ centered on beam 1, $\mathbf{k_{i}}+\mathbf{k_{j}}=(k_{1}+k_{2})\, \mathbf{\hat{e_{z}}}$, and the cuts have a thickness $10^{-2}~k_{\textrm{rec}}$ ($1.5 \times 10^{-1}~k_{\textrm{rec}}$) along $z$ ($x$ and $y$). The bunching, due to the correlation between the two beams, has a longitudinal (transverse) width $\sigma_{c,z}=1.8\times 10^{-2}~k_{\textrm{rec}}$ ($\sigma_{c,y}=1.6\times 10^{-1}~k_{\textrm{rec}}$). \\
(c,d) Cuts along ($y$,$z$) of the integrated, normalized local correlation function of beam 1, $g^{(2)}_{L}(\mathbf{\Delta k} ) = \int \mathbf{dk_{i}} \, g_{L}^{(2)}(\mathbf{k_{i}}, \mathbf{k_{i}}+ \mathbf{\Delta k})$. The integration region is the same as for the cross correlation, and the cuts have a thickness $2.5 \times 10^{-3}~k_{\textrm{rec}}$ ($0.1~k_{\textrm{rec}}$) along $z$ ($x$ and $y$). The bunching, due to HBT effect, has a longitudinal (transverse) width $\sigma_{l,z}=3.7\times 10^{-3}~k_{\textrm{rec}}$ ($\sigma_{l,y}=1.3\times 10^{-1}~k_{\textrm{rec}}$). Cuts along $x$ (not shown here) have same widths and amplitudes as cuts along $y$. These correlation functions are calculated using 850 experimental realisations, with $k_{0}= -0.65~k_{\textrm{rec}}$, a lattice depth $V_{0}=0.725~E_{\textrm{rec}}$ and a lattice duration $T_{L}=2$~ms. In all plots, the horizontal error bars indicate the bin size and the vertical ones correspond to the statistical $1\sigma$ uncertainties. The solid lines are Gaussian fits to the data from which we extract the correlation widths.
}
\end{center}
\end{figure}

In the following, we focus on the two beams. 
Using them for quantum atom optics experiments or for interferometry will require recombining them. 
It is therefore crucial to know the width of their correlation. 
From the 3D-momentum distribution $n(\mathbf{k})$, we computed the normalized second-order cross correlation function, 

\begin{equation}
\label{g2eq}
g^{(2)}_{C}(\mathbf{k},\mathbf{k'})=\frac{\left \langle n(\mathbf{k})\, n(\mathbf{k'}) \right \rangle }{\left \langle n(\mathbf{k}) \right \rangle \, \left \langle n(\mathbf{k'}) \right \rangle }
\end{equation}

\noindent where $\mathbf{k}$ belongs to beam 1 and $\mathbf{k'}$ to beam 2. 
The BEC is not exactly at rest in the optical trap, but exhibits shot-to-shot momentum fluctuations on the order of $10^{-2}\,k_{\textrm{rec}}$.
We correct for these fluctuations by recentering separately the single shot momentum distributions $n(\mathbf{k})$ around $k_ {1}$ and $k_ {2}$, using the shift obtained from Gaussian fits to the peak at $k_ {1}$ and to the diffraction peak at $k_ {0}+ 2 k_ {\textrm{rec}}$. 
This correlation function exhibits a peak for $k_{z} \simeq k_{1}$ and $k'_{z} \simeq k_{2}$ [Figs.~\ref{corFig}(a) and \ref{corFig}(b)]. 
The presence of this peak indicates that the two atomic beams are indeed correlated. 

We wish to determine the number of modes present in each beam, and how many of these modes are correlated.
We therefore examine the \textit{local} second-order correlation function of a single beam, $g^{(2)}_{L}(\mathbf{k},\mathbf{k'})$, which is obtained as in Eq.~(\ref{g2eq}) but with both $\mathbf{k}$ and $\mathbf{k'}$ belonging to beam 1. 
This correlation function, plotted in Figs.~\ref{corFig}(c) and \ref{corFig}(d), exhibits bunching for $k'_{z} \simeq k_{z} \simeq k_{1}$, due to density fluctuations [as in the Hanbury Brown-Twiss (HBT) effect \cite{Molmer2008}]. 
Similar bunching is observed at $k_{2}$. 
If we suppose that the widths of the local correlation define the size of a single mode, 
we can compare them to those of the density (longitudinal rms: $4\times10^{-2}~k_{\textrm{rec}}$, transverse rms: $4 \times 10^{-1}~k_{\textrm{rec}}$). 
We see that about 10 longitudinal and 3 transverse modes are populated. Thus the mode population is, roughly, $70~$atoms/mode. 
For comparison, in the case of free-space four-wave mixing \cite{Jaskula2010}, starting from a similar initial BEC, $10^{5}$ modes were populated, with only about $0.02~$atoms/mode. 

It appears in Fig.~\ref{corFig} that, while in the transverse direction, the cross and local correlations have similar widths [Figs.~\ref{corFig}(a) and \ref{corFig}(c)], the cross correlation is 5 times broader than the local one along the vertical axis [Figs.~\ref{corFig}(b) and \ref{corFig}(d)]: each mode is correlated with several modes of the other beam. 
If one uses two such beams as inputs to a beam splitter, this broadening amounts to a loss of coherence, and the interference contrast would be reduced.
We emphasize that the observed widths may be broadened by other effects, and so their numerical ratio is not exactly equal to the number of correlated modes.
For the local correlation, we estimate that the finite vertical resolution of the microchannel
plate detector contributes notably to the observed width.  
This resolution comes about because the surface which defines the atom arrival time is not flat but consists of tilted channels which intercept the atoms at different heights.  
The width shown in Fig.~\ref{corFig}(d) is consistent with this interpretation.
For the cross correlation, the observed width is broadened by the fact that the vertical source size is not negligible \cite{Zin}. Note also that the limited coherence of the initial quasi-BEC plays a role in the cross correlation width \cite{Zin}.

\begin{figure}
\begin{center}
\includegraphics[width=7cm]{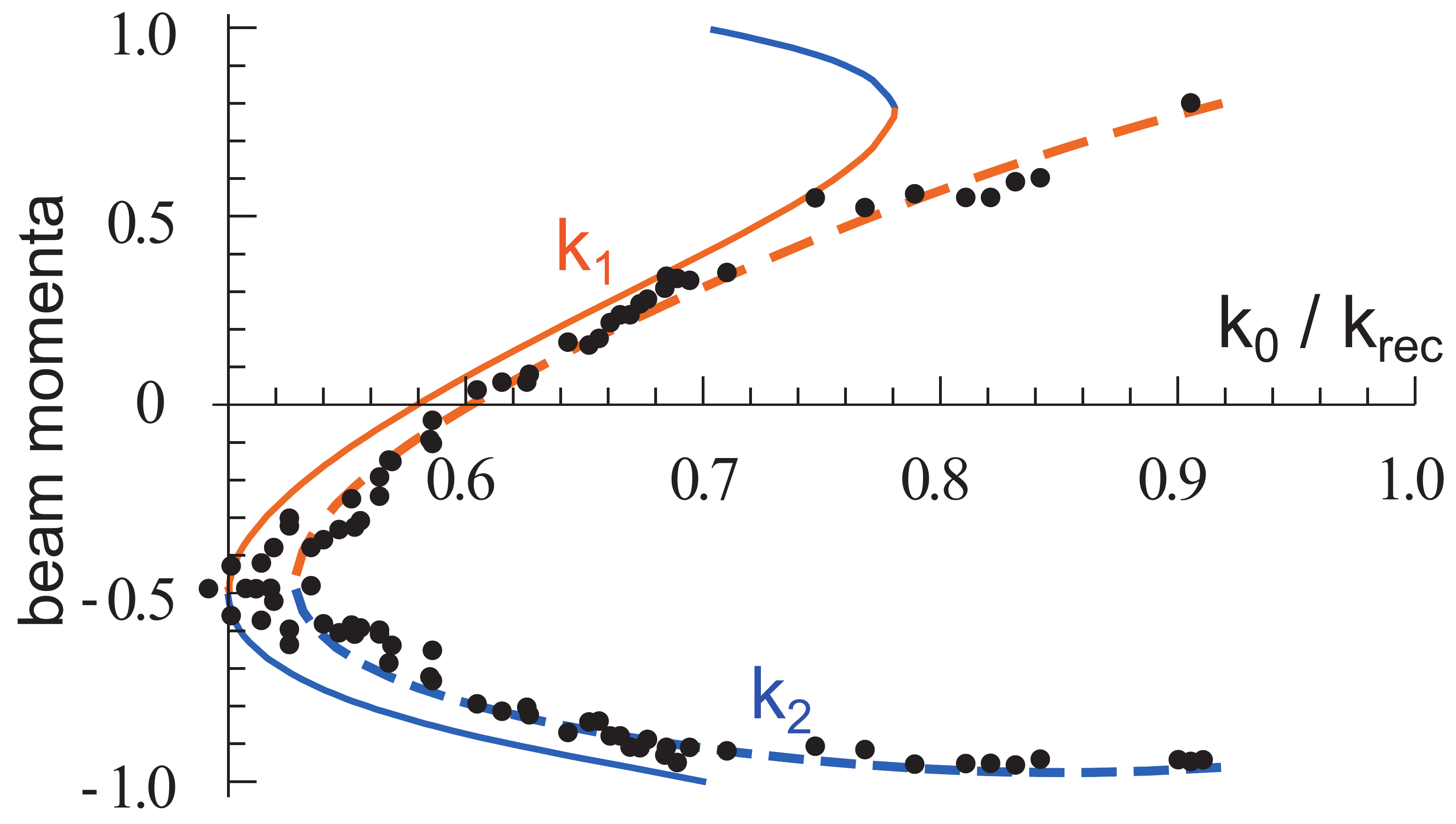}
\caption{\label{tuneFig}Measured mean momenta $k_{1}$ and $k_{2}$ of the beams (black dots, in units of $k_{\textrm{rec}}$) as a function of $k_{0}$ (initial BEC momentum in the lattice frame) for a depth $V_{0}=1.05~E_{\textrm{rec}}$ and a duration $T_{L}=1.5$~ms of the lattice. 
The solid line shows the phase-matching curve expected without interactions, while the dashed line includes the mean field [see Eq.~(\ref{MeanField})].}
\end{center}
\end{figure}

The use of an optical lattice permits control over the output beam momenta.
Changing the detuning $\delta \nu$ between the lattice beams results in varying the value of $k_{0}$. 
In Fig.~\ref{tuneFig}, we plot the mean vertical momenta $k_{1}$ and $k_{2}$ of both beams, measured for different $k_{0}$, 
as well as the expectation (solid line) based on the phase-matching conditions illustrated in Fig.~\ref{princFig}(a). 
We obtain a fair agreement over a large range, even though the solid line presents a small shift in comparison to the data points and does not reproduce the observed shape for high values of $k_{0}$. However, as already observed for four-wave mixing in free space \cite{Krachmalnicoff2010}, phase-matching conditions can be influenced by mean-field effects. 
A simple correction to the phase-matching curve is found just by 
adding the mean field to the energy conservation condition: Since the two atoms of a scattered pair are distinguishable from the atoms of the initial BEC, the mean-field energy experienced by each of them is not $g n_{0}$ (with $g=4 \pi \hbar^{2} a /m$, $a$ and $m$ the scattering length and the mass of He$^{*}$ and $n_{0}\simeq 10^{13}$ atoms/cm$^{3}$ the BEC's density), but $2 g n_{0}$, so that the energy conservation condition reads:
\begin{equation}
2 E(k_{0}) + 2 g n_{0} =  E(k_{1}) + E(k_{2}) + 4 g n_{0}
\label{MeanField}
\end{equation}
where the energy $E(k)$ is given by the dispersion relation in the first Bloch band of the lattice without interaction. 
As seen in Fig.~\ref{tuneFig} (dashed line), this correction leads to very good agreement with the experimental data, and accounts for the shift of the phase-matching curve and the change of its shape. 
A more exact calculation of the phase-matching conditions, inspired by Ref.~\cite{Wu2001}, confirms the accuracy of Eq.~(\ref{MeanField}) in our experimental conditions and will be given in Ref.~\cite{Ruaudel}.\par

\begin{figure}
\begin{center}
\includegraphics[width=7cm]{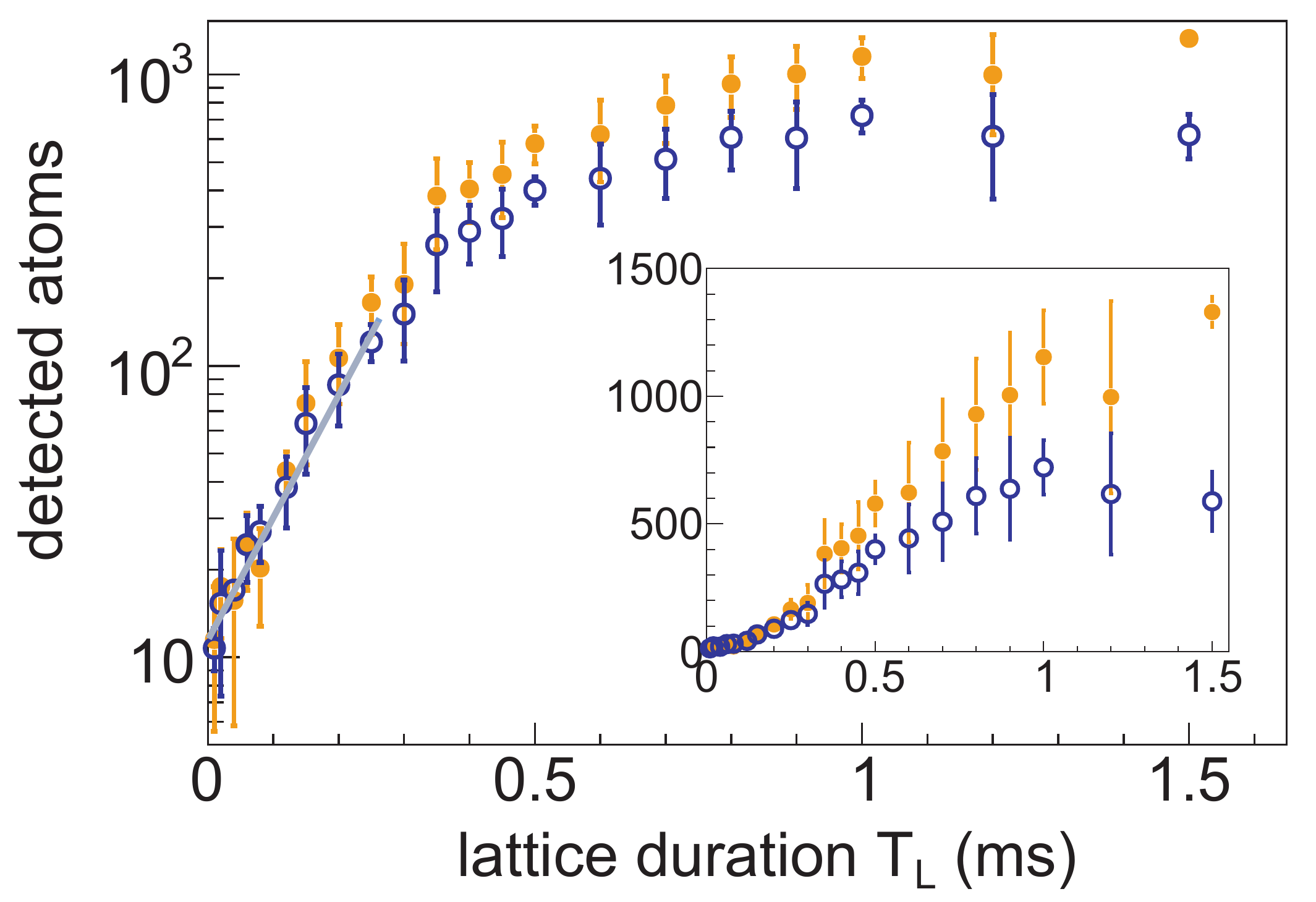}
\caption{\label{timeFig}(Color online) Dependence of the population of beam 1 (orange filled circles) and beam 2 (blue open circles) on the lattice duration $T_{L}$ for $k_{0}=-0.67~k_{\textrm{rec}}$ and for a lattice depth $V_{0}~=~1.05~E_{\textrm{rec}}$. 
The gray line is an exponential fit of the detected population in beam 2 for $T_{L}<0.3~$ms, which gives a time constant of 0.1~ms and an offset of 11.5 detected atoms. 
This offset is due to the small thermal part of the source cloud with quasimomenta $k_{1}$ and $k_{2}$. 
For a lower lattice depth, as for the data of Fig.~\ref{corFig}, the temporal evolution is a few times slower \cite{Wu2001}.  Inset: same data with linear scale.
}
\end{center}
\end{figure}

Another degree of freedom results from the fact that pair creation only takes place while the lattice is on.
We can thus tune the beam populations with the lattice duration $T_{L}$. 
In the example of Fig.~\ref{timeFig} these populations increase exponentially with $T_{L}$ during a few hundreds of $\mu$s, and then reach a plateau. 
This saturation could be explained by several mechanisms such as the decrease of spatial overlap between condensate and scattered beams \cite{Campbell2006}, multimode effects \cite{Bucker2012} and secondary scatterings from the beams. 
Condensate depletion is at most about 20~\%, and should be of little importance in the saturation. 
For small $T_{L}$, there is no discernable population difference between both beams. 
By contrast, we observe that at large $T_L$ the population of beam 1 is almost twice that of beam 2, a phenomenon also noticed in Ref.~\cite{Campbell2006}. 
This may be due to $k_2$ being in a dynamically unstable region while atoms with quasimomentum $k_1$ can only undergo secondary scattering to excited transverse modes.

\begin{figure}
\begin{center}
\includegraphics[width=7cm]{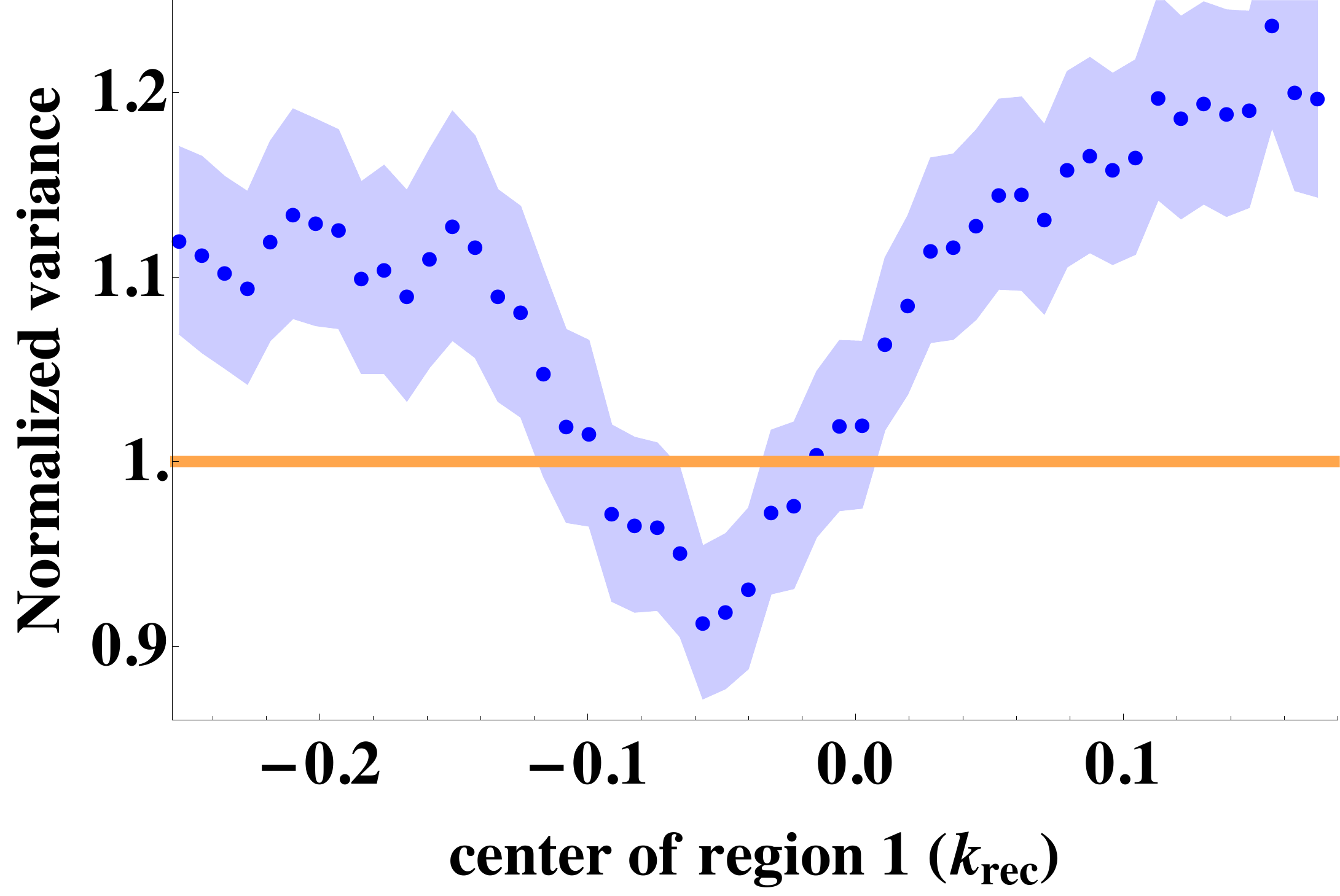}
\caption{\label{fig6}(Color online) Normalized variance of atom number difference between two regions selected close to beams 1 and 2. The data is the same as that of Fig.~\ref{corFig}. Regions are vertical cylinders of radius $2.5\times 10^{-2}~k_{\textrm{rec}}$ and height $8.5\times 10^{-2}~k_{\textrm{rec}}$. They are centered on the two beams in the transverse plane. Along the vertical axis, the center momentum (in the lattice frame) of region 1 is scanned, whereas region 2 is fixed. A variance below unity indicates sub-Poissonian fluctuations.}
\end{center}
\end{figure}

At intermediate $T_L$, we observe negligible losses due to secondary scattering \emph{and} high mode population (around 60 atoms per mode at $T_L=0.2$~ms in the example of Fig.\ref{timeFig}). 
The resulting beams should contain strongly correlated pairs.
In an attempt to verify a nonclassical correlation, we examined atom number difference between the two beams.
By selecting two regions around the centers of the two beams, we do indeed observe a sub-Poissonian number difference \cite{Jaskula2010,Bucker2011}, as shown
in Fig.~\ref{fig6}. The observed variance is consistent with that observed in Ref.~\cite{Jaskula2010}, and is limited in large part by the quantum efficiency of the detector. Other features of the variance are puzzling, however. 
First the minimun of the dip in the variance occurs when the center of region 1 is shifted by 0.1 $k_\textrm{rec}$ with respect to the center of the density distribution in beam 1. 
Second, in the transverse plane, the size of the regions over which the variance is reduced is nearly an order of magnitude smaller than the transverse width of the correlation function. We plan to investigate these effects in future experiments.

To conclude, we have demonstrated an efficient process for the production of correlated atom pairs. 
We have control over both the final momenta and the intensity of the correlated beams. We characterize the width of the correlation in momentum and find evidence of sub-Poissonian fluctuations of population difference. 
This source should be useful in multiple particle interference experiments both in the regime of well-isolated pairs \cite{Kofler2012} and in the regime of large occupation numbers \cite{Ferris2009}.

We thank K. M{\o}lmer, M. Ebner, J. Schmiedmayer and P. Zin for useful discussions. J.R. is supported by the DGA, R.L. by the FCT scholarship SFRH/BD/74352/2010, and support for the experimental work comes from the IFRAF program, the Triangle de la Physique, the ANR Grant ProQuP and ERC Grant 267 775 Quantatop.

\end{document}